\documentclass[12pt,english,a4paper]{article}
\usepackage[T1]{fontenc}
\usepackage[latin9]{inputenc}
\usepackage{amsmath}
\usepackage{graphicx}

\makeatletter

\usepackage{float}

\usepackage{epsfig}

\makeatother

\usepackage{babel}

\begin{document}

\makeatother

\title{\textbf{\large Note on triangle anomaly with improved momentum cutoff}}

\author{G. Cynolter and E. Lendvai}

\date{Theoretical Physics Research Group of Hungarian Academy of Sciences,
Eötvös University, Budapest, 1117 Pázmány Péter sétány 1/A, Hungary }
\maketitle
\begin{abstract}
A Lorentz and gauge symmetry preserving regularization method has
been proposed recently in 4 dimension based on Euclidean momentum
cutoff. It is shown that the triangle anomaly can be calculated unambiguously
with this new improved cutoff. The anticommutator of $\gamma^{5}$
and $\gamma^{\mu}$ multiplied by five $\gamma$ is proportional to
terms that do not vanish under a divergent loop-momentum integral,
but cancel otherwise.
\end{abstract}

\section{Introduction}

In quantum field theories higher order perturbative calculations need
symmetry preserving regularization. The most popular and very effective
regularization is dimensional regularization (DREG) \cite{dreg}.
DREG respects Lorentz and gauge symmetries, but as it modifies the
number of dimensions (at least in the loops) it is not directly applicable
to chiral theories like the standard model or to supersymmetric theories.
Continuation of $\gamma_{5}$ to dimensions $d\neq4$ goes with a
$\gamma_{5}$ not anticommuting with the extra elements of gamma matrices,
and it leads to {}``spurious anomalies'', see \cite{collins,spuri,jegerl,korner},
and references therein. The renormalizability can only be maintained
by imposing in each order the Ward-Takahashi or Slavnov-Taylor identities
manually and this process makes the calculations more complicated.
Pauli-Villars regularizations is straightforward, but the subtracted
propagators are not physical and there are problems at higher loop
calculations, see \cite{osipov} and \cite{brigi}. 4-dimensional
momentum cutoff is a simple regularizations scheme, but in its original
form it badly violates symmetries. There were proposals recently to
modify the calculation with momentum cutoff to respect Lorentz and
gauge symmetries \cite{uj,Olesz,liao,gu,wu1,rosten}. In this paper
we investigate the improved momentum cutoff proposed in \cite{uj},
successfully applied to a non-renormalizable theory \cite{fcmlambda}.
The loop integrals using this new regularization are invariant against
the shift of the loop momentum, therefore the usual derivation of
the ABJ triangle anomaly fails in this case (can not pick up a finite
term shifting the linear divergence). In what follows we show that
the proper handling of the trace of $\gamma_{5}$ and six gamma matrices
provides the correct anomaly, the $\left\{ \gamma_{5},\gamma_{\mu}\right\} $
anticommutator does not vanish in special cases under divergent loop
integrals. 

The rest of the paper is organized as follows. In section 2. the improved
momentum cutoff is summarized, in section 3. the triangle anomaly
is discussed, the paper is closed with conclusion and an appendix
with useful integrals.

\section{Improved momentum cutoff}

A new regularization is proposed in \cite{uj} based on 4 dimensional
momentum cutoff to evaluate 1-loop divergent integrals. The loop integrals
are calculated as follows. First the loop momentum ($k$) integral
is Wick rotated (to $k_{E}$), with Feynman parameter(s) the denominators
are combined, then the order of Feynman parameter and the momentum
integrals are changed. After that the loop momentum ($k_{E}\rightarrow l_{E}$)
is shifted to have a spherically symmetric denominator. 

The main observation was that contraction with $g_{\mu\nu}$ not necessarily
commutes with loop-integration in divergent cases. Therefore the substitution
of \begin{equation}
k_{\mu}k_{\nu}\rightarrow\frac{1}{4}g_{\mu\nu}k^{2}\label{eq:negyed}\end{equation}
is not acceptable under divergent integrals%
\footnote{The metric tensor is denoted by $g_{\mu\nu}$ both in Minkowski and
Euclidean space. %
}. The usual factor $1/4$ is resulted by tracing both sides under
a loop integral, which cannot be proven a valid step for Wick-rotated
divergent integrals in Minkowski space. It is better to define the
integrals with free Lorentz indices using physical consistency conditions,
like gauge invariance or freedom of momentum routing. Based on the
diagrammatical proof of gauge invariance it can be shown that the
two conditions are related and both are in connection with the requirement
of vanishing surface terms. It is shown in \cite{uj} that instead
of \eqref{eq:negyed} the general identification of the cutoff regulated
integrals \begin{equation}
\int_{\Lambda\: reg}d^{4}l_{E}\frac{l_{E\mu}l_{E\nu}}{\left(l_{E}^{2}+m^{2}\right)^{n+1}}:=\frac{1}{2n}g_{\mu\nu}\int_{\Lambda\: reg}d^{4}l_{E}\frac{1}{\left(l_{E}^{2}+m^{2}\right)^{n}},\ \ \ \ \ n=1,2,...\label{eq:idn}\end{equation}
will satisfy the Ward-Takahashi identities and gauge invariance at
1-loop. It differs from \eqref{eq:negyed} only in case of divergent
integrals, for finite cases both substitutions give the same results
(the surface terms vanish). It is shown in \cite{uj} that this definition
is robust, differently organized calculations of the 1-loop functions
agree with each other using \eqref{eq:idn} and disagree using \eqref{eq:negyed}.
For more than two free indices the consistency conditions give ($n=2,3,...$)\begin{equation}
\int_{\Lambda\: reg}d^{4}l_{E}\frac{l_{E\alpha}l_{E\beta}l_{E\mu}l_{E\rho}}{\left(l_{E}^{2}+m^{2}\right)^{n+1}}:=\frac{1}{4n(n-1)}\int_{\Lambda\: reg}d^{4}l_{E}\frac{g_{\alpha\beta}g_{\mu\rho}+g_{\alpha\mu}g_{\beta\rho}+g_{\alpha\rho}g_{\beta\mu}}{\left(l_{E}^{2}+m^{2}\right)^{n-1}}.\label{eq:idn4}\end{equation}
For 6 and more free indices there are appropriate rules, or \eqref{eq:idn}
can be used recursively. Finally the integrals are evaluated with
a Euclidean momentum cutoff.

In this method the terms with numerators proportional to the loop
momentum are all defined by symmetry. Odd number of $l_{E}$'s give
zero as usual, but the integral of even number of $l_{E}$ are defined
by \eqref{eq:idn} and \eqref{eq:idn4}, this guarantees that the
symmetries are not violated. The calculation is performed in 4 dimensions,
the finite terms are equivalent with DREG, and the method identifies
quadratic divergencies while gauge and Lorentz symmetries are respected.
We stress that the method treats differently momenta with free Lorentz
indices ($k_{\mu}k_{\nu}$) and indices summed up ($k^{2}$), the
order of tracing and performing the regulated integral cannot be changed
similarly to DREG.

The shift of the loop momentum does not generate surface terms, just
as in DREG, but this property would make the triangle anomaly disappear
in a naive calculation. In the next section we show that the new method
provides a well defined result for the famous triangle anomaly.

\section{Triangle anomaly}

\begin{figure}
\begin{centering}
\includegraphics[width=11cm]{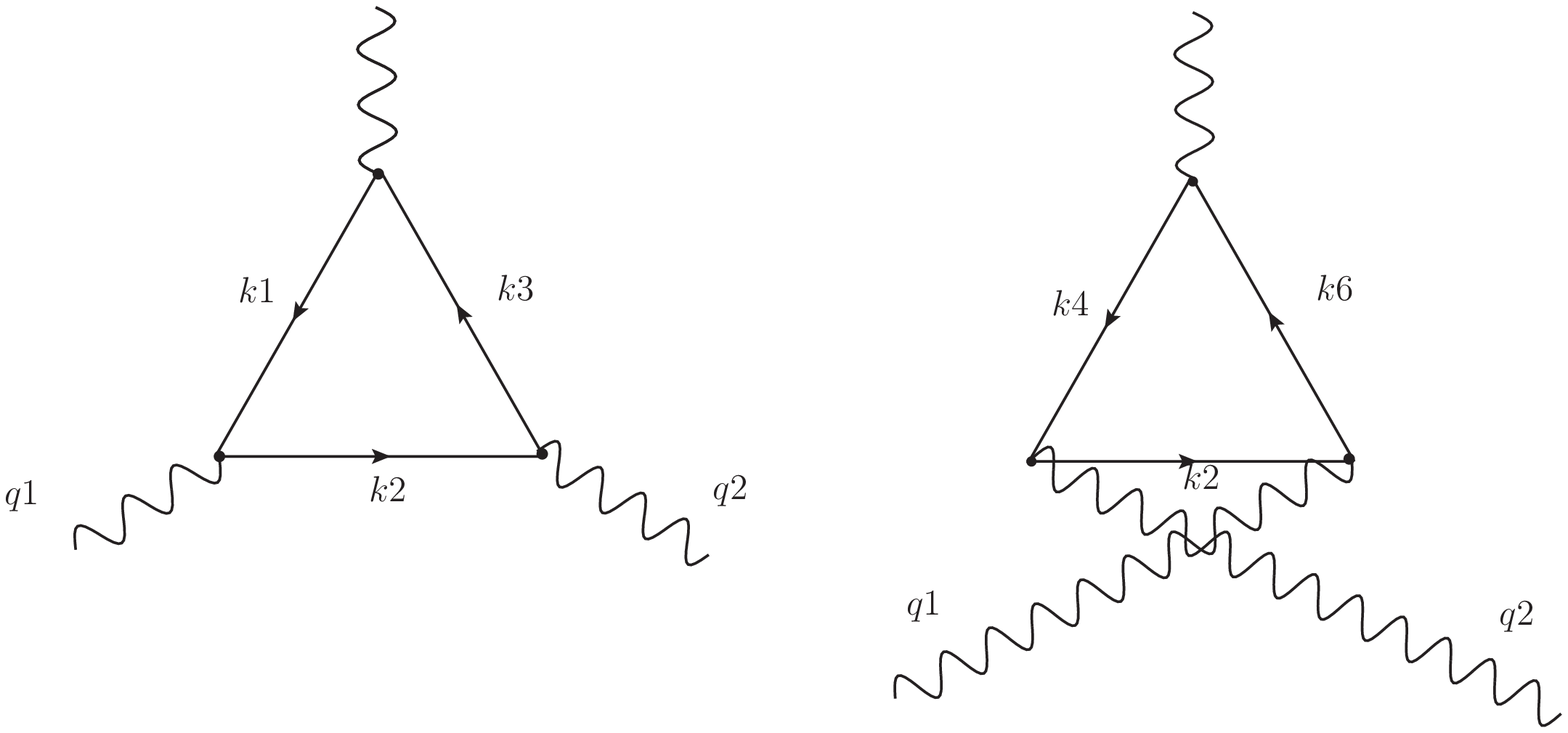}
\par\end{centering}

\centering{}\caption{Feynman graphs contributing to the triangle anomaly, $k_{2}=k$, $k_{1}=k-q_{1}$,
$k_{3}=k+q_{2}$, $k_{4}=k-q_{2}$ and $k_{6}=k+q_{1}$. }

\end{figure}
In the present method the triangle anomaly has to be recalculated.
Consider the 1-loop triangle graph on the left on Fig. 1.\begin{equation}
T_{1}^{\mu\nu\rho}=e^{2}\int\frac{d^{4}k}{(2\pi)^{4}}Tr\left(\gamma^{5}\frac{\not k-\not q_{1}+m}{\left(k-q_{1}\right)^{2}-m^{2}}\gamma^{\mu}\frac{\not k+m}{k^{2}-m^{2}}\gamma^{\nu}\frac{\not k+\not q_{2}+m}{\left(k+q_{2}\right)^{2}-m^{2}}\gamma^{\rho}\right).\phantom{qq}\label{eq:triang1}\end{equation}
The amplitude of the crossed graph $T_{2}^{\mu\nu\rho}$ is similar
with $(q_{1},\,\mu)$ and $(q_{2},\,\nu)$ interchanged ($T^{\mu\nu\rho}=T_{1}^{\mu\nu\rho}+T_{2}^{\mu\nu\rho}$).
The Ward identities require\begin{eqnarray}
q_{1\mu}T^{\mu\nu\rho} & = & 0,\label{eq:a1}\\
q_{2\nu}T^{\mu\nu\rho} & = & 0,\label{eq:a2}\\
-(q_{1}+q_{2})_{\rho}T^{\mu\nu\rho} & = & 2mT^{5\mu\nu},\label{eq:a3}\end{eqnarray}
where $T^{5\mu\nu}$ corresponds to the same graphs with a pseudoscalar
current instead of the axialvector one. There is a formal proof of
(\ref{eq:a3}). Replace

\begin{equation}
-(q_{1}+q_{2})_{\rho}\gamma^{\rho}\gamma^{5}=-\left(\not k+\not q_{2}-m\right)\gamma^{5}+\left(\not k-\not q_{1}-m\right)\gamma^{5}.\label{eq:alak}\end{equation}
The first term combines with the numerator of the last term in \eqref{eq:triang1}
and cancels the denominator. If \begin{equation}
\left\{ \gamma^{\mu},\gamma^{5}\right\} =0\label{eq:AC}\end{equation}
assumed, then the second term in \eqref{eq:alak} is $-\left(\not k-\not q_{2}-m\right)\gamma^{5}=+\gamma^{5}\left(\not k-\not q_{2}-m\right)+2m\gamma^{5}$.
Here the first term cancels the adjacent fraction in \eqref{eq:triang1}
and the second term gives the right hand side of \eqref{eq:a3}. The
$\left(-(q_{1}+q_{2})_{\rho}T_{1}^{\mu\nu\rho}-2mT_{1}^{5\mu\nu}\right)$
difference is\begin{equation}
e^{2}\int\frac{d^{4}k}{(2\pi)^{4}}\left(\gamma^{5}\frac{\not k-\not q_{1}+m}{\left(k-q_{1}\right)^{2}-m^{2}}\gamma^{\mu}\frac{\not k+m}{k^{2}-m^{2}}\gamma^{\nu}+\gamma^{5}\gamma^{\mu}\frac{\not k+m}{k^{2}-m^{2}}\gamma^{\nu}\frac{\not k+\not q_{2}+m}{\left(k+q_{2}\right)^{2}-m^{2}}\right).\label{eq:diff}\end{equation}
Shifting $k\rightarrow k+q_{1}$ in the first term and moving $\gamma^{\mu}$
through $\gamma^{5}$ (using again \eqref{eq:AC}) to the back of
the second term we arrive to a formula that is totally antisymmetric
under the interchange of $(q_{1},\,\mu)$ and $(q_{2},\,\nu)$, and
thus adding the crossed graph ($T_{2}^{\mu\nu\rho}$) the result vanishes.
Similarly \eqref{eq:a1} and \eqref{eq:a2} can be proven but here
\eqref{eq:AC} is not needed to apply, because the terms leading to
cancellation are not separated by a factor of $\gamma^{5}$. The loop
momentum can be shifted, this is a fundamental property of the improved
momentum cutoff regularization.

However (\ref{eq:a1}-\ref{eq:a3}) cannot be all true. Pauli-Villars
regularization or careful simple momentum cutoff calculation identifies
a finite anomaly term when shifting the linearly divergent integral,
though there is still an ambiguity in connection with momentum routing
where to put the anomaly term in (\ref{eq:a1}-\ref{eq:a3}). At the
same time in improved momentum cutoff or DREG \eqref{eq:a1} and \eqref{eq:a2}
holds but the proof of \eqref{eq:a3} is false%
\footnote{Functional integral derivation of the anomaly shows that the Ward
identity corresponding to the axial vector current \eqref{eq:a3}
must be anomalous \cite{Fuji}. %
}, it relies additionally on \eqref{eq:AC}. This is the first sign
that the naive anticommutator \eqref{eq:AC} can not be used in all
situations.

The explicit calculation of the triangle diagram \eqref{eq:triang1}
is based on the evaluation of the trace of $\gamma^{5}$ with six
$\gamma$'s. There are various methods to calculate this trace with
superficially different terms at the end. The different results of
the trace can be transformed to each other using the Schouten identity,
a special form of it reads\begin{equation}
-k^{2}\epsilon_{\mu\nu\lambda\rho}+k^{\alpha}k_{\mu}\epsilon_{\alpha\nu\lambda\rho}+k_{\nu}k^{\alpha}\epsilon_{\mu\alpha\lambda\rho}+k_{\lambda}k^{\alpha}\epsilon_{\mu\nu\alpha\rho}+k_{\rho}k^{\alpha}\epsilon_{\mu\nu\lambda\alpha}=0.\label{eq:schouten}\end{equation}
In the present method this identity cannot be used for the loop momentum
($k$) of a divergent integral before applying the identifications
\eqref{eq:idn} or \eqref{eq:idn4}, because it would mix free Lorentz
indices and indices summed up, which must be evaluated in a different
way (DREG faces the same difficulty). After performing the identifications
\eqref{eq:idn} and \eqref{eq:idn4} the quadratic loop momenta factors
cancel with the denominators. The remaining formula contains the loop
momentum in the numerators at maximum linearly, the corresponding
Schouten identity can be applied. The root of the problem is that
in case of divergent integrals the totally antisymmetric tensor $\epsilon_{\mu\nu\lambda\rho}$
can not taken out of the integral, similarly to the case of $g_{\mu\nu}$
in the previous section. No problem emerges for finite integrals. 

The breakdown of the early application of the Schouten identity forces
us to choose one dedicated calculation of the trace. The trace is
calculated not using the anticommutator \eqref{eq:AC}, just \begin{equation}
\left\{ \gamma_{\mu},\gamma_{\nu}\right\} =2g_{\mu\nu}\,,\label{eq:clifford}\end{equation}
and general properties of the trace. The unambiguous result is \begin{eqnarray}
 &  & \frac{1}{4}\mathrm{Tr}\left[\gamma_{5}\gamma_{\alpha}\gamma_{\mu}\gamma_{\beta}\gamma_{\nu}\gamma_{\rho}\gamma_{\lambda}\right]=\epsilon_{\alpha\mu\beta\nu}g_{\rho\lambda}-\epsilon_{\alpha\mu\beta\rho}g_{\nu\lambda}+\epsilon_{\alpha\mu\nu\rho}g_{\beta\lambda}-\epsilon_{\alpha\beta\nu\rho}g_{\mu\lambda+}\nonumber \\
 &  & +\epsilon_{\mu\beta\nu\rho}g_{\alpha\lambda}-\epsilon_{\lambda\alpha\mu\beta}g_{\rho\nu}+\epsilon_{\lambda\alpha\mu\nu}g_{\rho\beta}-\epsilon_{\lambda\alpha\beta\nu}g_{\rho\mu}+\epsilon_{\lambda\mu\beta\nu}g_{\rho\alpha}-\epsilon_{\lambda\rho\alpha\mu}g_{\nu\beta}+\nonumber \\
 &  & +\epsilon_{\lambda\rho\alpha\beta}g_{\nu\mu}-\epsilon_{\lambda\rho\mu\beta}g_{\nu\alpha}+\epsilon_{\lambda\rho\nu\alpha}g_{\mu\beta}-\epsilon_{\lambda\rho\nu\mu}g_{\alpha\beta}+\epsilon_{\lambda\rho\nu\beta}g_{\alpha\mu}.\label{eq:tr6}\end{eqnarray}
It reflects the complete Lorentz structure of the $\gamma$ matrices
in the trace. This choice of the trace appeared in earlier papers
without detailed argumentations \cite{polon},\cite{mawu}. All different
calculations of the trace are in agreement with each other and with
\eqref{eq:tr6} if \eqref{eq:AC} is modified. $\gamma_{5}$ and $\gamma_{\mu}$
does not always anticommute (rather the anticommutator picks up terms
proportional to the Schouten identity.) Explicitly, the following
definition will eliminate all the ambiguities burdening the calculation
of the trace of $\gamma_{5}$ and six $\gamma$'s \begin{eqnarray}
\mathrm{Tr}\left[\left\{ \gamma_{\rho},\gamma_{5}\right\} \gamma_{\lambda}\gamma_{\alpha}\gamma_{\mu}\gamma_{\beta}\gamma_{\nu}\right]=\mathrm{2Tr}\left[g_{\nu\rho}\gamma_{5}\gamma_{\lambda}\gamma_{\alpha}\gamma_{\mu}\gamma_{\beta}-g_{\beta\rho}\gamma_{5}\gamma_{\lambda}\gamma_{\alpha}\gamma_{\mu}\gamma_{\nu}+\right.\nonumber \\
\left.+g_{\mu\rho}\gamma_{5}\gamma_{\lambda}\gamma_{\alpha}\gamma_{\beta}\gamma_{\nu}-g_{\alpha\rho}\gamma_{5}\gamma_{\lambda}\gamma_{\mu}\gamma_{\beta}\gamma_{\nu}+g_{\lambda\rho}\gamma_{5}\gamma_{\alpha}\gamma_{\mu}\gamma_{\beta}\gamma_{\nu}\right].\label{eq:trg5}\end{eqnarray}
The above anticommutator is defined only under the trace. \eqref{eq:trg5}
can be understood as the $\left\{ \gamma_{5},\gamma_{\rho}\right\} $
anticommutator is defined by picking up all the terms when moving
$\gamma_{\rho}$ all the way round through all the other $\gamma$'s.
Evaluating the trace the right hand side is proportional to Schouten
identities. Under a divergent loop integral it will not vanish in
the present method (nor in DREG). The nontrivial anticommutator contributes
to the triangle anomaly but vanishes in nondivergent cases and for
less $\gamma$'s. The amplitude of the triangle diagrams can be calculated
with the definition of the trace \eqref{eq:tr6} and the rules \eqref{eq:idn},
\eqref{eq:idn4}. Finally we arrive at the extra anomaly term in \eqref{eq:a3}.

In what follows we calculate directly the anomaly term missing in
\eqref{eq:a3}. We use \eqref{eq:alak} and move $\left(\not k-\not q_{1}-m\right)$
from the back to the front in \eqref{eq:triang1} using \eqref{eq:clifford}.
Without this trick the trace of six $\gamma$'s and $\gamma^{5}$
should have been calculated uisng \eqref{eq:tr6}, which is consistent
with non-anticommuting $\gamma^{5}$ in this special case, see \eqref{eq:trg5}.
\begin{eqnarray}
-(q_{1}+q_{2})_{\rho}T_{1}^{\mu\nu\rho} & = & e^{2}\int\frac{d^{4}k}{(2\pi)^{4}}\mathrm{Tr}\left[\phantom{\gamma^{\mu}\frac{1}{\not k+m}}\right.\nonumber \\
 &  & -\gamma^{5}\frac{1}{\not k-\not q_{1}-m}\gamma^{\mu}\frac{1}{\not k-m}\gamma^{\nu}+\gamma^{5}\gamma^{\mu}\frac{1}{\not k+m}\gamma^{\nu}\frac{1}{\not k+\not q_{2}+m}+\phantom{++}\nonumber \\
 &  & +2\gamma^{5}\frac{1}{\not k-\not q_{1}-m}\gamma^{\mu}\frac{1}{\not k-m}\gamma^{\nu}\frac{(k-q_{1})(k+q_{2})}{(k+q_{2})^{2}-m^{2}}-\nonumber \\
 &  & -2\gamma^{5}\frac{1}{\not k-\not q_{1}-m}\gamma^{\mu}\frac{1}{\not k-m}\gamma^{\nu}\frac{(k^{\nu}-q_{1}^{\nu})}{\not k+\not q_{2}+m}+\nonumber \\
 &  & +2\gamma^{5}\frac{1}{\not k-\not q_{1}-m}\gamma^{\mu}\gamma^{\nu}\frac{1}{\not k+\not q_{2}+m}\frac{(k-q_{1})(k)}{k^{2}-m^{2}}-\nonumber \\
 &  & \left.-2\gamma^{5}\frac{1}{\not k-\not q_{1}-m}\frac{1}{\not k+m}\gamma^{\nu}\frac{(k^{\mu}-q_{1}^{\mu})}{\not k+\not q_{2}+m}\right].\label{eq:15}\end{eqnarray}
With algebraic manipulations using the antisymmetry of the trace including
$\gamma_{5}$ and four $\gamma$'s we can group the terms\begin{eqnarray}
-(q_{1}+q_{2})_{\rho}T_{1}^{\mu\nu\rho} &  & =e^{2}\int\frac{d^{4}k}{(2\pi)^{4}}\mathrm{Tr\gamma^{5}\left[\frac{\not k\not q_{1}\gamma^{\mu}\gamma^{\nu}}{N_{1}N_{2}}+\frac{\not k\not q_{2}\gamma^{\mu}\gamma^{\nu}}{N_{2}N_{3}}+2m^{2}\frac{\not q_{1}\not q_{2}\gamma^{\mu}\gamma^{\nu}}{N_{1}N_{2}N_{3}}+\right.\phantom{11}}\nonumber \\
 &  & +\frac{2}{N_{1}N_{2}N_{3}}\left\{ -\not q_{1}\not q_{2}\gamma^{\mu}\gamma^{\nu}\cdot k^{2}+\not k\not q_{2}\gamma^{\mu}\gamma^{\nu}\cdot kq_{1}+\phantom{\frac{2}{2}}\right.\nonumber \\
 &  & \left.\phantom{\frac{2}{2}}+\not q_{1}\not k\gamma^{\mu}\gamma^{\nu}\cdot kq_{2}+\not q_{1}\not q_{2}\not k\gamma^{\nu}\, k^{\mu}+\not q_{1}\not q_{2}\gamma^{\mu}\not k\, k^{\nu}\right\} +\nonumber \\
 &  & +\frac{2}{N_{1}N_{2}N_{3}}\left(+\not\not k\not q_{1}\gamma^{\mu}\gamma^{\nu}\cdot q_{1}q_{2}-\not q_{2}\not q_{1}\gamma^{\mu}\gamma^{\nu}\cdot kq_{1}-\phantom{\frac{2}{2}}\right.\nonumber \\
 &  & \left.\left.\phantom{\frac{2}{2}}-\not\not k\not q_{2}\gamma^{\mu}\gamma^{\nu}\cdot q_{1}^{2}-\not k\not q_{1}\not q_{2}\gamma^{\nu}\, q_{1}^{\mu}-\not k\not q_{1}\gamma^{\mu}\not q_{2}\, q_{1}^{\nu}\right)\right],\label{eq:long}\end{eqnarray}
where $N_{1}=\left((k-q_{1})^{2}-m^{2}\right)$, $N_{2}=\left(k^{2}-m^{2}\right)$
and $N_{3}=\left((k+q_{2})^{2}-m^{2}\right)$. The first two terms
vanish after performing the trace and the integral (they are proportional
to $\epsilon(\mu,\nu,q_{1},q_{1})\equiv\epsilon_{\mu\nu\alpha\beta}q_{1}^{\alpha}q_{1}^{\beta}$
and $\epsilon(\mu,\nu,q_{2},q_{2})$ respectively). The third one
gives $2m$ times the pseudoscalar amplitude $T^{5\mu\nu}=T_{1}^{5\mu\nu}+T_{2}^{5\mu\nu}$,\begin{equation}
T_{1}^{5\mu\nu}=-m\epsilon(\mu,\nu,q_{1},q_{2})e^{2}\int\frac{d^{4}k}{(2\pi)^{4}}\left[\frac{1}{N_{1}N_{2}N_{3}}\right],\label{eq:pvv1}\end{equation}
we get $T_{2}^{5\mu\nu}$ interchanging $(q_{1},\mu)\leftrightarrow(q_{2},\nu)$
in the integrand.

The last five terms in \eqref{eq:long} contain \textit{one factor}
of the loop momentum ($k$) and after tracing vanish by the Schouten
identity, the loop integration does not spoil the cancellation. The
contribution of the one but last five terms in the curly bracket does
not vanish. It contains \textit{two factor} of the loop momentum,
and it is proportional to Schouten identity \eqref{eq:schouten} broken
under the divergent loop integral. Calculating it with the improved
momentum cutoff of Section 2 using the formulas of the Appendix (or
with DREG) we get the anomaly term.

\begin{equation}
-(q_{1}+q_{2})_{\rho}T^{\mu\nu\rho}=2mT^{5\mu\nu}-i\frac{e^{2}}{2\pi^{2}}\epsilon^{\mu\nu\alpha\beta}q_{1\alpha}q_{2\beta}.\label{eq:a03A}\end{equation}
In the case of the naive substitution \eqref{eq:negyed} the Schouten
identity \eqref{eq:schouten} is satisfied, the curly bracket vanishes.
(In that case with simple momentum cutoff the anomaly term originates
from shifting the linearly divergent first two terms in \eqref{eq:long},
but the result depends on momentum routing.) The presented method
identifies without ambiguity the value of the anomaly in the axial-vector
current and leaves the vector currents anomaly free without any further
assumptions.

\section{Conclusion}

We have investigated the triangle anomaly within the 4 dimensional
improved momentum cutoff framework. This regularization respects gauge
and Lorentz symmetries by construction, the loop-integrals are invariant
under the shift of the loop momentum. This property spoils the usual
derivation of the ABJ anomaly in the presence of a cutoff. We have
chosen to calculate the trace corresponding to the triangle graphs
of Fig. 1. ($\gamma_{5}$ and six $\gamma$'s) and the Ward identity
\eqref{eq:a03A} ($\gamma_{5}$ and four $\gamma$'s) only using the
standard anticommutators of the $\gamma$ matrices \eqref{eq:clifford}.
It turns out that different evaluation of the trace will agree with
each other if and only if $\gamma^{5}$ does not always anticommute
with $\gamma^{\mu}$, rather $\left\{ \gamma^{\mu},\gamma^{5}\right\} $
picks up terms proportional to the Schouten identity \eqref{eq:trg5}
if it is multiplied with five more $\gamma$'s under the trace. Multiplying
the $\left\{ \gamma^{\mu},\gamma^{5}\right\} $ anticommutator with
three $\gamma$'s, it vanishes as $\mathrm{Tr}(\gamma^{5}\gamma_{\alpha}\gamma_{\beta}\gamma_{\mu}\gamma_{\nu})$
is unambiguous. The right hand side of \eqref{eq:trg5} is only non-vanishing
if it is under a divergent loop momentum integral, where at least
two factors of the loop momentum is involved in the identity. The
nontrivial properties of $\gamma^{5}$ and $\gamma$'s first appear
in field theory in the divergent triangle diagram. 

Traces involving $\gamma^{5}$ and even number of $\gamma$'s can
be calculated in the same manner avoiding the anticommutation of $\gamma^{\mu}$and
$\gamma^{5}$. First the order of $\gamma^{\nu}$'s are reversed applying
\eqref{eq:clifford} then using the cyclicity of the trace we get
back the original trace in the reversed order, the difference gives
the trace twice. This way the $\left\{ \gamma^{\mu},\gamma^{5}\right\} $
anticommutator can be also defined, it will not vanish generally.
If it is multiplied with (2n-1) $\gamma$'s it is equal to the sum
of (2n-1) trace involving $\gamma^{5}$ and (2n-2) $\gamma$'s, see
\eqref{eq:trg5}. It is well known that the general properties of
the trace and $\left\{ \gamma^{\mu},\gamma^{5}\right\} =0$ are in
conflict with each other, this led to the 't Hooft- Veltman scheme
\cite{dreg,collins}. Our proposal similarly modifies $\left\{ \gamma^{\mu},\gamma^{5}\right\} $
but works in four dimensions and the modifications come into action
only under divergent loop integrals involving enough $\gamma$ matrices.
There were attempts to keep $\left\{ \gamma^{\mu},\gamma^{5}\right\} =0$,
but then the cyclicity of the trace was lost \cite{korner}.

We stress that our method works in the four physical dimensions. We
have shown that the vector currents are conserved and the axial vector
current is anomalous, and no ambiguity appears. As a future work the
improved momentum cutoff could be applied to higher loops or non-abelian
gauge theories, it is promising as the implicit momentum regularization
fulfilling similar consistency conditions successful at more than
one loops \cite{nemes2}. The strength of the improved momentum cutoff
method is that it can be used in theories with quadratic divergencies
important for example in gauge theories including gravitational interactions
\cite{toms}.

\appendix

\section{Useful integrals}

In this appendix we list the divergent integrals used for the triangle
anomaly calculated by the new regularization. $\Delta$ can be any
loop momentum $(k)$ independent expression depending on the Feynman
$x$ parameter, external momenta, etc., e.g. $\Delta(x,q_{i},m).$
The integration is understood for Euclidean momenta with absolute
value below $\Lambda$.

The integral \eqref{eq:A1} is just given for comparison, it is calculated
with a simple momentum cutoff. In \eqref{eq:A2} with the standard
\eqref{eq:negyed} substitution one would get a constant $-\frac{3}{2}$
instead of $-1$ \cite{uj}.

\begin{eqnarray}
\int_{\left|k_{E}\right|\leq\Lambda}\frac{d^{4}k}{i(2\pi)^{4}}\frac{1}{\left(k^{2}-\Delta^{2}\right)^{2}} & \!\!=\!\! & \frac{1}{(4\pi)^{2}}\left(\ln\left(\frac{\Lambda^{2}+\Delta^{2}}{\Delta^{2}}\right)+\frac{\Delta^{2}}{\Lambda^{2}+\Delta^{2}}-1\right).\label{eq:A1}\\
\int_{\left|k_{E}\right|\leq\Lambda}\frac{d^{4}k}{i(2\pi)^{4}}\frac{k_{\mu}k_{\nu}}{\left(k^{2}-\Delta^{2}\right)^{3}} & \!\!=\!\! & \frac{1}{(4\pi)^{2}}\frac{g_{\mu\nu}}{4}\left(\ln\left(\frac{\Lambda^{2}+\Delta^{2}}{\Delta^{2}}\right)+\frac{\Delta^{2}}{\Lambda^{2}+\Delta^{2}}-1\right).\label{eq:A2}\end{eqnarray}

\end{document}